\documentclass[%
twocolumn,aps,pra,
superscriptaddress,
 amsmath,amssymb]{revtex4-1}

\usepackage{xcolor}

\usepackage{graphicx}
\usepackage{braket}
\begin{document}

\title{Illuminating the bulk-boundary correspondence of a non-Hermitian stub lattice with Majorana stars}

\author{James Bartlett}
\affiliation{%
 Department of Physics and Astronomy, George Mason University, Fairfax, Virginia 22030, USA
}%
\author{Haiping Hu}
\affiliation{%
 Department of Physics and Astronomy, George Mason University, Fairfax, Virginia 22030, USA
}%
\affiliation{%
Department of Physics and Astronomy, University of Pittsburgh, Pittsburgh, Pennsylvania 15260, USA}%
\author{Erhai Zhao}
\affiliation{%
 Department of Physics and Astronomy, George Mason University, Fairfax, Virginia 22030, USA
}%


\begin{abstract}
Topological characterization of non-Hermitian band structures demands more than a straightforward 
generalization of the Hermitian cases.
Even for one-dimensional tight-binding models with nonreciprocal hopping, the appearance of point gaps and the
skin effect leads to the breakdown of the usual bulk-boundary correspondence. Luckily, the correspondence 
can be resurrected by introducing a winding number for the generalized Brillouin zone for systems with an
even number of bands and chiral symmetry. Here, we analyze the topological phases of 
a nonreciprocal hopping model on the stub 
lattice, where one of the three bands remains flat. Due to the lack of chiral symmetry, the biorthogonal Zak phase is
no longer quantized, invalidating the winding number as a topological index. Instead, we show that a $Z_2$
invariant can be defined from Majorana's stellar representation of the eigenstates on the
Bloch sphere. The parity of the total azimuthal winding of the entire Majorana constellation correctly predicts
the appearance of edge states between the bulk gaps. 
We further show that the system is not a square-root topological insulator, despite the fact
that its parent Hamiltonian can be block diagonalized and related to a sawtooth lattice model. 
The analysis presented here may be generalized to understand 
other non-Hermitian systems with multiple bands.
\end{abstract}

\maketitle

\section{\label{sec:intro}Introduction}

Non-Hermitian (NH) Hamiltonians have long been adopted to describe a wide range of open or nonequilibrium quantum systems.
The experimental realization of $\mathcal{PT}$-symmetric systems in quantum optics has renewed
the interest in NH lattice systems, especially their topological properties \cite{Ruter10}. From a theoretical point of view,
NH systems are interesting because they host a number of unique phenomena such as exceptional points in the spectrum \cite{Berry04,Heiss12,Miri19} and the NH skin effect \cite{Yao18,Kunst18,Lee16,Xiong18,Yoko19,Yang20,Zhang21,Ghatak20,Helbig20,Xiao20,Hofmann20,Borgnia20}, where an extensive number of eigenstates are accumulated at the boundaries. These properties promise new technological applications including topological lasing \cite{Peng14,Brand14,Harari18,Bandres18} and enhanced quantum sensing \cite{Budich20,Bao21}. 

Substantial theoretical progress has been made to systematically classify and characterize the topological band structures of NH Bloch Hamiltonians.
The initial schemes were based on gap dichotomy, i.e. by differentiating point gaps from line gaps \cite{Gong18,Kawa19,Kawa19_2,Zhou19,Liu19,Liu19_2}. Later on the more general cases of separable bands \cite{Shen18} were considered, leading to their homotopy classification using braid groups \cite{Wojcik20,Li21} and knots \cite{Hu21} in one dimension (1D). A major obstacle in developing a full NH band theory is the sensitivity of the spectrum to the boundary conditions. Bulk topological invariants defined for real quasimomentum $k$ and periodic boundary condition are usually insufficient to describe the excitations at the open boundary which may include the skin modes. This problem is well recognized in one dimension. For example, for the NH Su-Schrieffer-Heeger (SSH) model \cite{PhysRevLett.42.1698,Yao18,Lieu18,Kunst18,Lee16}, the phase diagram predicted by the winding number does not agree with the appearance of edge states, signaling the breakdown of the usual bulk-edge correspondence. 

One way to recover the correspondence is to generalize the Bloch band theory to allow
the quasimomentum $k$ to be complex and analytically continue the Hamiltonian $H(k)$ to $H(\beta=e^{ik})$. Then one can define the so-called generalized Brillouin zone (GBZ) $C_\beta$, a closed curve on the complex $\beta$ plane. Taking the GBZ curve $C_\beta$ as the base manifold, a winding number can be defined analogous to the Hermitian SSH model, which correctly predicts the emergence of edge states \cite{Yao18,Lieu18,Kunst18,Lee16}. 
The definition of the winding number over $C_\beta$ requires the presence of chiral symmetry, which is common for bipartite lattice models with nearest neighbor hopping and  an even number of bands. For 1D lattice models that lack the chiral symmetry, it remains unclear whether it is viable, or how, to construct a proper NH topological invariant.

To address this open question, in this paper we study a 1D NH tight-binding model on the stub lattice with three bands  (see Fig. 1). It turns out that, perhaps counterintuitively, the topological characterization of the NH stub model is rather nontrivial. The Zak phase \cite{Zak89} accumulated when transversing the curve $C_\beta$  is not quantized, for the model lacks chiral symmetry. Nonetheless, it features robust edge states in certain parameter regimes to indicate a topologically nontrivial bulk. We present two ways to characterize the bulk topology, the first via the Majorana star representation and the second through the decomposition of the squared Hamiltonian $H^2(\beta)$. Both methods are capable of recovering the generalized (non-Bloch) bulk-boundary correspondence, and their results are consistent with each other. 

A secondary motivation to examine the NH stub lattice is to elucidate the interplay of flat bands \cite{leykam2018artificial} and the skin effect. It is well known that the stub lattice features a completely flat band at zero energy \cite{real2017flat}. Another well known lattice that possesses a flat band, the Lieb lattice in 2D \cite{PhysRevLett.62.1201}, can be viewed as the stub lattice stacked together. The lack of dispersion means the kinetic energy is frustrated. In fact, the degenerate states within the flat band are compact-localized \cite{flach2014detangling} in real space, i.e., the corresponding Wannier functions have a compact support and vanish beyond a finite cluster size. Then it is natural to expect them to resist the NH skin effect. Recall that in the simplest case of the NH skin effect, e.g. in the Hatano-Nelson model \cite{hatano1996localization}, all eigenstates are localized to one edge due to nonreciprocal hopping. Such a scenario seems improbable for the flat band.
Since compact localized states are potentially useful for optical applications, such as the diffraction-free propagation of light \cite{Vicencio14,Vicencio15,Mukherjee15,Dai20} and enhanced light-matter interaction by generating slow light \cite{Krauss07,Baba08,Li08,Schulz17},
it is worthwhile to investigate their NH skin effect. 

This paper is organized as follows. In Sec. \ref{sec:model}, we introduce the NH stub lattice model and discuss its bulk spectra. We also analyze its band topology from the perspective of knot theory \cite{Hu21}. Then in Sec. III, we present the edge spectrum and a systematic analysis of the NH skin effect by comparing the localization properties of the continuum bands measured by the inverse participation ratio.  Comparing the bulk and edge spectrum points to the failure of the usual bulk-boundary correspondence.
To restore the correspondence, we introduce the notion of the generalized Brillouin zone, continuum band and bi-orthogonal Zak phase in Sec. IV.
These setups enable us to define in Sec. \ref{sec:majorana} a topological invariant based on Majorana's stellar representation \cite{Liu14,Teo20,Xu20}. We show the invariant yields the correct prediction of the edge states.  Section \ref{sec:sqrt} is devoted to a simplified picture of the azimuthal winding through the parent Hamiltonian \cite{Arkinstall17,Ezawa20}. We conclude in Sec. \ref{sec:sum} by discussing open questions and possible experimental realizations of the stub model.

\section{\label{sec:model}The non-Hermitian stub model}

Our starting point is a Hermitian tight-binding model on the stub lattice, schematically shown in Fig. 1 with $\kappa=0$. 
Each unit cell contains three sites, $a$, $b$, and $c$. The hopping amplitudes $t_{1,2,3}$ are real.
The bulk energy spectrum has three bands, one of which is completely flat at zero energy,
\begin{equation} \label{hermi}
E(k)=0, \pm \sqrt{t_1^2 +t_2^2 +t_3^2 +2 t_1t_2\cos k}.
\end{equation}
Here, $k\in [-\pi,\pi]$ is the quasimomentum and we have set the unit cell size to be one. The existence of the flat band 
can be understood as follows. The whole lattice can be partitioned into two sublattices: sublattice A consisting of the $a$ and $c$ sites, 
and sublattice B of all the $b$ sites. Particles only hop between the sublattices, so the Hamiltonian has a sublattice symmetry.
Define projector $P_A$ ($P_B$) for the A (B) sublattice, then the operator $\Gamma=P_A-P_B$ anticommutes with $H$,
$\Gamma H \Gamma = -H$ with $\Gamma^2=1$. Then it is straightforward to show the number of zero modes 
$N_0=\mathrm{Tr} \Gamma$ \cite{guzman2021geometry}. 
The trace of $\Gamma$ has a simple interpretation in real space, it is the imbalance of the number of
sites within the two sublattices, $\mathrm{Tr} \Gamma=N_A-N_B$. Thus $N_0=N_A-N_B$ which is a general and well-known
result. Applying it to the stub lattice, we recover the result above that at each $k$, there is exactly one zero mode. 
In other words, we have a flat band at zero energy.

\begin{figure}[htpb!]
\includegraphics[width=0.4\textwidth]{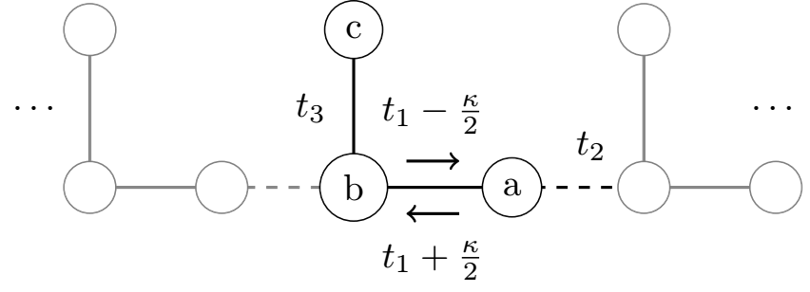}
\caption{\label{fig:stub} Schematic of the non-Hermitian stub lattice. Each unit cell consists of three sites labeled by $a$, $b$, and $c$, respectively.
The hopping parameters $t_{1,2,3}$ and $\kappa$ are all real. For finite $\kappa$, the horizontal intracell hopping between $a$ and $b$ is nonreciprocal,
so the tight-binding Hamiltonian becomes non-Hermitian. We take $t_2=1$ as the energy unit.}
\end{figure}

\begin{figure}[htpb!]
\includegraphics[width=0.45\textwidth]{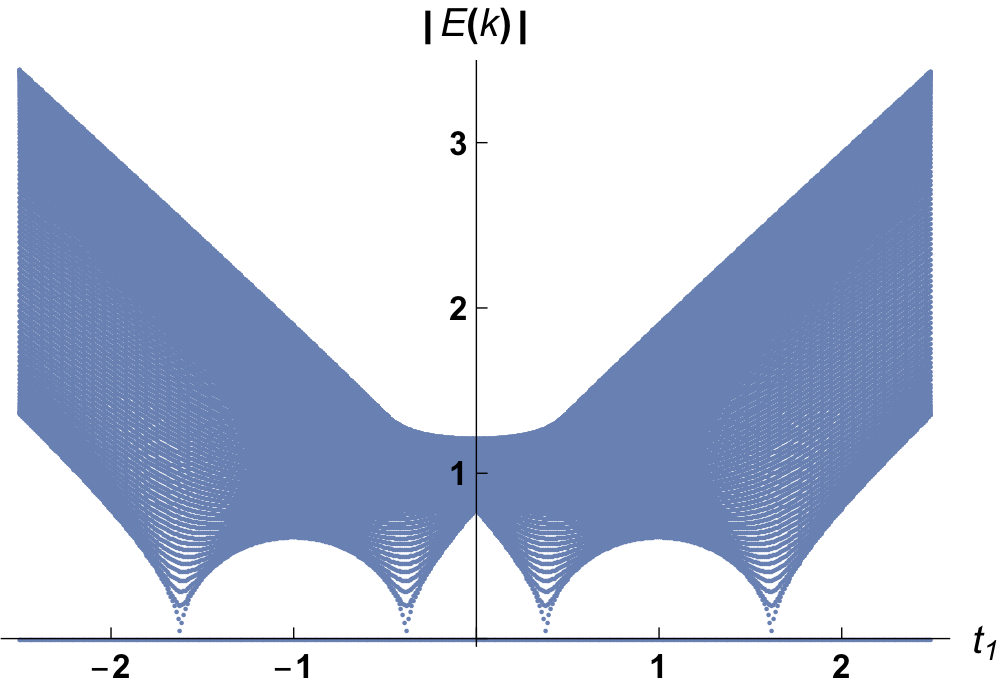}
\caption{\label{fig:bulk} Bulk energy spectrum of the NH stub model with parameters $t_{2}=1,t_{3}=0.25,\kappa=4/3$. The magnitude of $E(k)$ plotted
for varying $t_{1}$ shows a flat band at zero energy, and two gap closing points at $t_L$=0.382 and $t_H=1.62$.}
\end{figure}

We generalize the stub lattice model by allowing the intracell hopping between the $a$ and $b$ sites to be nonreciprocal.
The hopping asymmetry is characterized by a parameter $\kappa$ as depicted in Fig. \ref{fig:stub}. 
The resultant tight-binding Hamiltonian in second quantized form is 
\begin{eqnarray}
H =\underset{n}{\sum}&&\big[ (t_{1}+\frac{\kappa}{2} )b_{n}^{\dagger}a_{n}+ (t_{1}-\frac{\kappa}{2} )a_{n}^{\dagger}b_{n}+t_{3}b_{n}^{\dagger}c_{n}\nonumber\\
&&+t_{3}c_{n}^{\dagger}b_{n}
	+t_{2}b_{n+1}^{\dagger}a_{n}+t_{2}a_{n}^{\dagger}b_{n+1}\big].
\label{eq:HamReal}
\end{eqnarray}
Here $a_{n}^{\dagger}$ creates a particle at the $a$ site of the $n$-th unit cell, and similarly for $b_n^\dagger$ and $c_n^\dagger$.
We assume $t_{1,2,3}$ and $\kappa$ are real, and set $t_2=1$ unless specified otherwise. 
In momentum space, the Bloch Hamiltonian is a $3\times 3$ matrix
\begin{equation}
H(k)=\left[\begin{array}{ccc}
0 & (t_{1}-\frac{\kappa}{2})+t_{2}e^{ik} & 0\\
(t_{1}+\frac{\kappa}{2})+t_{2}e^{-ik} & 0 & t_{3}\\
0 & t_{3} & 0
\end{array}\right]
\label{eq:HamBloch}.
\end{equation}
This non-Hermitian model does not have $\mathcal{PT}$ symmetry, so its eigenenergies are in general complex,
\begin{align}
&E(k)=0, \pm \sqrt{\alpha_k+it_2\kappa\sin k}, \\
&\alpha_k = t_1^2 +t_2^2 +t_3^2-\kappa^2/4 +2 t_1t_2\cos k.
\end{align}
The bulk energy spectrum is illustrated in Fig. \ref{fig:bulk}, where the magnitude of $E(k)$ is plotted against $t_1$.
There is a flat band at zero energy, just as in the Hermitian case, and the energy gap closes at two critical values,
$t_1=t_L$ and $t_H$ (it is sufficient to focus on $t_1>0$). The values of $t_{L,H}$ can be easily obtained by solving
$\alpha_{k=\pi}=0$, a quadratic equation for $t_1$ yielding two roots. For the parameters given in Fig. \ref{fig:bulk}, 
$t_L=0.382$ and $t_H=1.62$.

\begin{figure}[htpb!]
\includegraphics[width=0.4\textwidth]{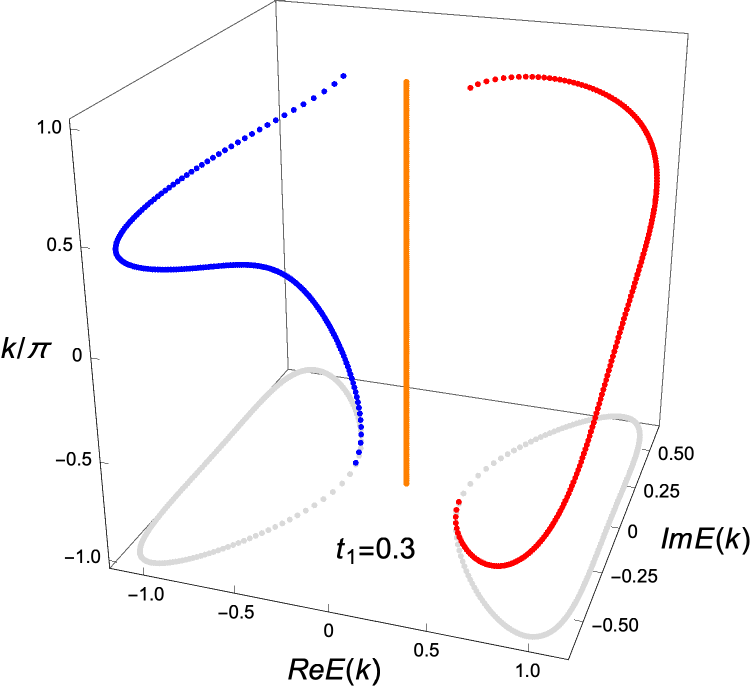}
\includegraphics[width=0.4\textwidth]{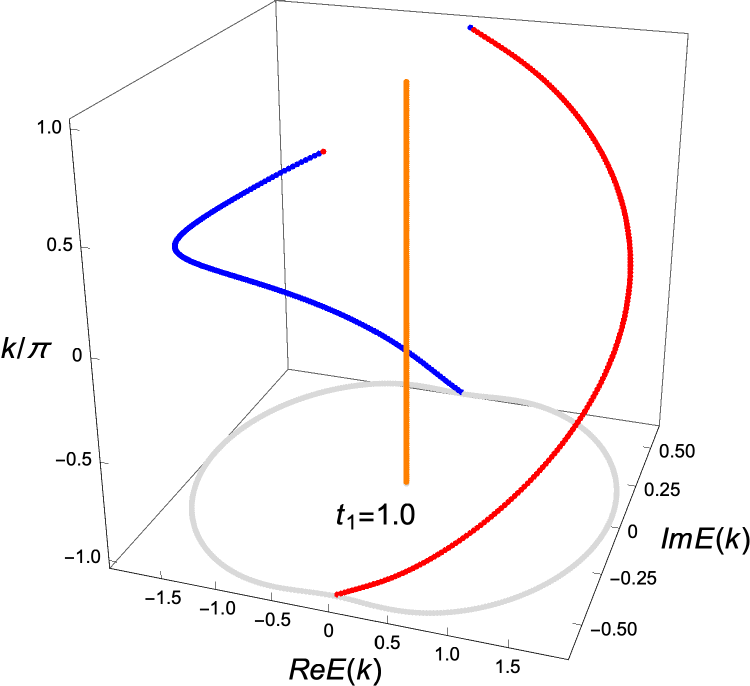}
\caption{\label{fig:braid} Two bulk phases identified from the braiding pattern of the eigenenergy strings: the unlink phase (upper panel, for $t_1=0.3$)
and the unknot phase (lower panel, for $t_1=1.0$). Parameters $t_2$, $t_3$, and $\kappa$ are the same as Fig. \ref{fig:bulk}.}
\end{figure}

These gap-closing points mark the transition between two topologically distinct phases. To see this, it is best to 
plot the eigenenergy string in the space spanned by (Re$E$, Im$E$, $k$) as shown in Fig. \ref{fig:braid}.
For $t_1\in [0,t_L)$ and $t_1>t_H$, e.g. $t_1=0.3$ in the upper panel, there is no braiding between the two nonflat bands (in blue and red respectively). 
Projecting the spectrum on the complex energy plane (grey curves in Fig. \ref{fig:braid}), we see that each 
nonflat band forms a closed curve. There is no linking between the two curves, so we call it the unlink phase.
Note that for a reference energy $E_p$ inside either of the closed curves, the system has a point gap. 
In comparison, for $t_1\in(t_L,t_H)$ such as $t_1=1.0$ in the lower panel, the two energy strings braid once.
Since the Brillouin zone is periodic, during the evolution of $k$ from $-\pi$ to $\pi$, the ending 
point ($k=\pi$) of the blue band becomes the starting point ($k= -\pi$) of the red band. 
When projected on the complex energy plane, the two bands join each other to form a single curve, i.e., a trivial knot or unknot.
Note that the unknot phase also has a point gap.
The unlink and unknot phase are topologically distinct. It is impossible to continuously vary one into the other while
keeping the bands separated. In Ref. \cite{Hu21}, two of us showed that phase transitions between two
phases characterized by distinct knots/links occur at exceptional points. This can be verified numerically at $t_{L}$ and $t_H$. 

\section{\label{sec:skin}Edge states and skin effect}

\begin{figure}[htpb!]
\includegraphics[width=0.4\textwidth]{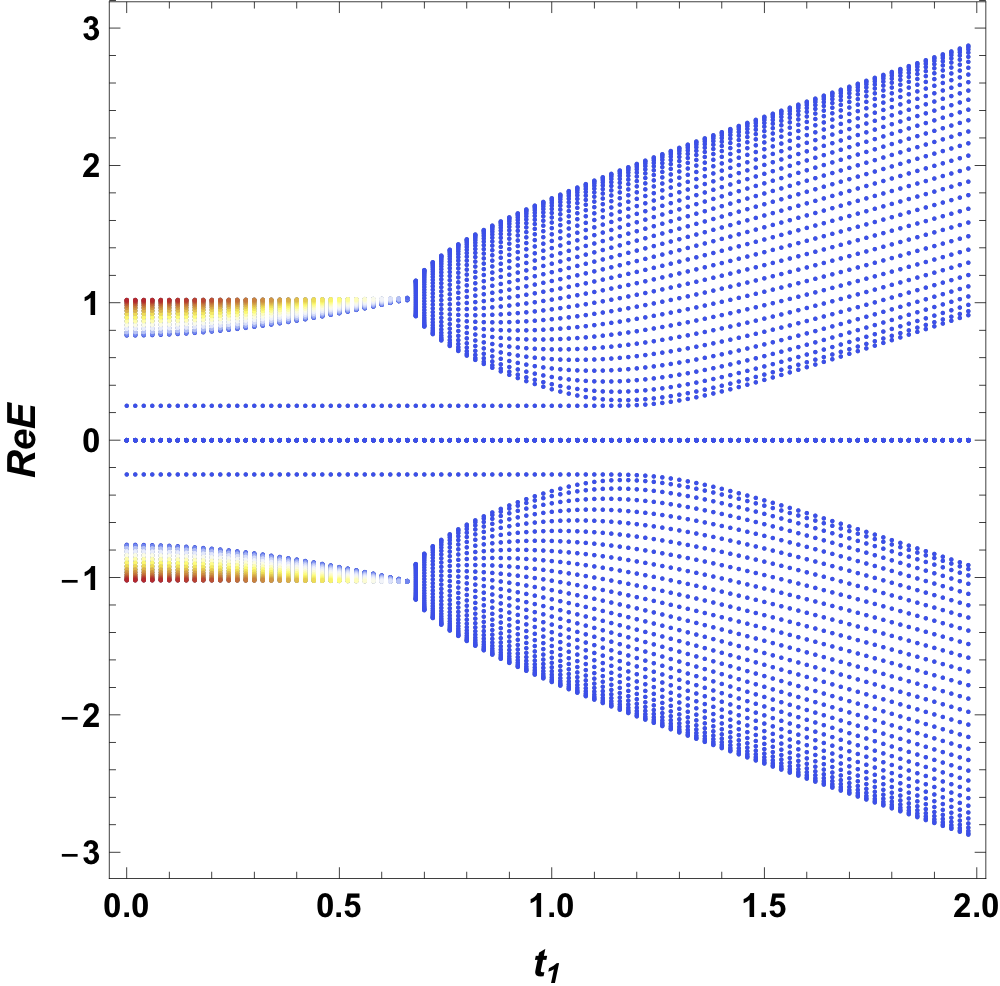}
\caption{\label{fig:chain} The energy spectrum of a finite chain of the nonreciprocal stub model
with parameters $t_{2}=1,t_{3}=0.25,\kappa=4/3$, $L=30$. The imaginary part of the energy eigenvalues is
indicated by color, with real energy shown in blue. The edge states living inside the bulk gap
are at $E_{edge}=\pm t_3$ and they persist up to $t_1=t_c=1.2$.}
\end{figure}

Next we show that these phase transition points for the bulk band structure do not coincide with 
where the edge states change qualitatively. Consider a finite chain of $L$ unit cells terminating at the edge as depicted in Fig. \ref{fig:stub}. 
An example of its energy spectrum is given in Fig. \ref{fig:chain} for $L=30$.
Here the vertical axis shows the real part of the energy Re$E$, while $|\mathrm{Im}E|$ is indicated by
color. In particular, all the points in blue represent real energy eigenvalues.
Comparing to the bulk spectrum with the same parameters shown in Fig. \ref{fig:bulk},
we see that the flat band at zero energy persists in the finite lattice. Moreover, 
edge states appear inside the bulk gap at
\begin{equation}
E_{edge} = \pm t_3
\end{equation}
for $|t_{1}|<t_c$. 
Note that the critical value $t_c$, in this case $t_c=1.2$, differs from $t_L$ or $t_H$ above. 
The discrepancy indicates the breakdown of bulk-edge correspondence, which is 
is well recognized in NH systems.

The edge states at energy $\pm t_3$ do not hinge on the model being non-Hermitian. It is
also present in the Hermitian limit $\kappa=0$. Its robustness is attested by its independence on $t_1$
(as long as $t_1<t_c$). Such independence also suggests that the origin of the edge state can be revealed by
considering the limit $t_1=\kappa=0$, i.e. when the chain is broken up into disjoint pieces.
Each piece is an elbow-shaped ``molecule'' consisting of sites $c_n$, $b_n$ and $a_{n-1}$ coupled by $t_3$ and $t_2$,
see Fig. \ref{fig:stub}. Then the eigenenergies of the whole chain are easy to enumerate. 
At the right edge $n=L$, site $a_L$ is dangling and not coupled to anything else to give
eigenenergy 0. At the left edge $n=1$, sites $b_1$ and $c_1$ are only coupled to each other by $t_3$,
which yields eigenenergy $\pm t_3$. For all the molecules in the middle, we have
eigenenergy 0, $\pm \sqrt{t_2^2+t_3^2}$. Thus, the edge states at $\pm t_3$ can be traced to 
the molecular states isolated at the edge in the limit of vanishing $t_1$ and $\kappa$.
The physical picture here is very analogous to the SSH model in the dimer limit. The difference is that
for the stub lattice, $E_{edge}$ is not at zero energy. Rather, it is repelled from the zero-energy flat bands
to reside inside the band gap.

Going away from the molecule limit by increasing $t_1$ while maintaining $\kappa=0$, the energy bands acquire 
the dispersion given in
Eq. \eqref{hermi}. We can determine when the edge states merge into the bulk bands as follows. The 
bottom of the upper band is $E(k=\pi)=\sqrt{(t_1-t_2)^2+t_{3}^2}$. Equating it to $E_{edge}=t_3$, 
we find that the merge occurs when $t_1=t_2$, which is consistent with the numerics. 
Applying the same reasoning to the NH stub model,
one might expect that the edge state would cease to exist when the bulk band bottom $\alpha_{k=\pi}=t_3$, 
i.e., when $t_1=(t_2- \kappa/2)$. However, this prediction based on the bulk spectrum is incorrect and does
not agree with Fig. \ref{fig:chain}. As we will show in Sec. IV below, the correct critical value for the NH model is 
\begin{align}
   & t_{c}=\sqrt{t_{2}^{2}+\left(\frac{\kappa}{2}\right)^{2}}&\quad\text{for}\quad\frac{\kappa}{2}<t_{1},\label{tca}\\
    &t_{c}=\sqrt{-t_{2}^{2}+\left(\frac{\kappa}{2}\right)^{2}}&\quad\text{for}\quad\frac{\kappa}{2}>t_{1}.\label{tcb}
\end{align}
The failure of the naive prediction for $t_c$ is another manifestation of the nontrivial bulk-edge correspondence 
in NH systems.  

\begin{figure}[htpb!]
\includegraphics[width=0.4\textwidth]{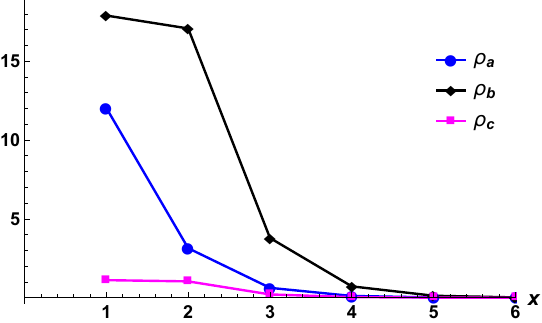}
\includegraphics[width=0.4\textwidth]{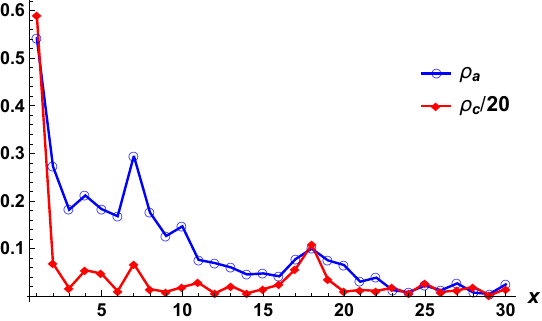}
\caption{\label{fig:skin} Non-Hermitian skin effect. Shown are the total probability densities $\rho_s(x)$, $s=a,b,c$, 
for the bulk band with Re$E<0$
(top) and the zero energy states (bottom, with $\rho_b$=0).  
$t_{2}=1,t_{3}=0.25,\kappa=4/3$, and $L=30$.}
\end{figure}

The presence of open boundaries drastically changes the wave functions of the three continuum bands.
For a given eigenenergy $E$, let $\psi_s(x)$ be its wave function (more precisely the
right eigenvector of $H$) at unit cell $x\in [1,L]$ and site
$s\in\{a,b,c\}$, and define the probability density $\rho_s(x)=|\psi_s(x)|^2$.  
For example, one finds that all $\psi_s(x)$ at energies with finite Re$E$ localize at the left boundary 
for $\kappa>0$. The upper panel of Fig. \ref{fig:skin} illustrates the total probability
$\rho_s$ of all states with Re$E<0$ (excluding the edge state at $-t_3$). They decay exponentially
into the bulk to exhibit NH skin effect. 
The flat band states at zero energy also gravitate toward the left edge, but the localization is far from 
complete and the decay is not exponential.
The lower panel of Fig. \ref{fig:skin} shows the total probability of all zero energy states as function of 
the unit cell index $x$. Note that zero energy states only live on the $a$ and $c$ sites. The resistance
to skin effect observed here is in accordance to the intuition based on the real space picture of flat band states
summarized in the introduction.

\begin{figure}[htpb!]
\includegraphics[width=0.4\textwidth]{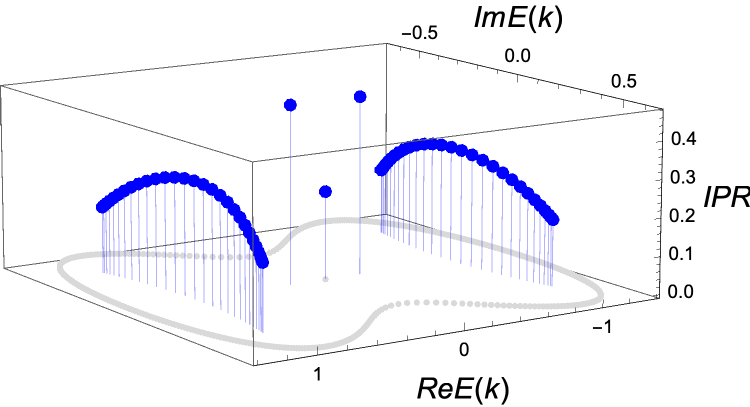}
\caption{\label{fig:ipr} The inverse participation ratio (IPR) of the eigenstates of
 a finite chain. The grey points are the eigenenergies of the bulk. 
 $t_{2}=1,t_{3}=0.25,\kappa=4/3$, $L=30$.}
\end{figure}

A more quantitative measure of the localization is provided by the inverse participation ratio,
$\mathrm{IPR}=\sum_{s,x} |\psi_s(x)|^4$ defined for a given energy. The value of IPR
approaches 1 for a state perfectly localized on one site, and order $1/L^2$ for a uniformly
delocalized state.
For degenerate states such as those at zero energy, the IPR is averaged within the degenerate subspace. Fig. \ref{fig:ipr} 
summarizes and compares the IPR for all the eigenenergies of a finite chain.
It is clear that the edge states at $\pm t_3$ are the most localized with the highest IPR.
Next are the continuum of states with finite Re$E$ that show the skin effect.
The least localized are the zero energy states. 
For reference, the energies of the bulk (with periodic boundary conditions) are shown in grey.
They form a closed ring to enclose all the open spectra on the complex energy plane.
The change from the grey to blue is an example of the sensitivity of the spectrum to the boundary conditions
in NH systems.

\section{\label{sec:gbzl}Generalized Brillouin zone}
The continuum of states for finite chains with open boundary conditions as shown in Figs. \ref{fig:chain} and \ref{fig:ipr} are not Bloch waves
with real wave number $k$. In order to describe them and account for the NH skin effect, an established procedure \cite{Yao18,Yoko19}
is to analytically continue $H(k)$ to $H(\beta)$ where $\beta=e^{ik}$ can take complex values away from the unit circle,
\begin{equation}
H\left(\beta\right)=\left [\begin{array}{ccc}
0 & \left(t_{1}-\frac{\kappa}{2}\right)+t_{2}\beta & 0\\
\left(t_{1}+\frac{\kappa}{2}\right)+t_{2}\beta^{-1} & 0 & t_{3}\\
0 & t_{3} & 0
\end{array}\right]
\label{eq:HamBeta}.
\end{equation}
Then, an eigenstate $\psi(x)\sim \beta^x$ can describe states localized at the boundaries. 
The value of $\beta$ is not arbitrary and must be chosen properly such that the energy spectrum of $H(\beta)$ given by
\begin{equation} \label{eign-H-beta}
\det\left[{H}(\beta)-E\right]=0
\end{equation}
matches that of a long chain in the limit $L\rightarrow \infty$. To this end,
the boundary conditions at $x=1,L$ must be met, and the spectrum of $H(\beta)$ must 
approach a union of continuum manifolds, known as continuum bands, in the limit of $L\rightarrow \infty$.
It was shown in Ref. \cite{Yoko19} that these requirements are met when the ``continuum condition'' is satisfied,
$|\beta_M|=|\beta_{M+1}|$, where $\beta_i$ with $i=1,2,..., 2M$ are the solutions to Eq. \eqref{eign-H-beta}, which
is an algebraic equation for $\beta$ for given $E$, with their magnitudes sorted in ascending order, $|\beta_1|\leq |\beta_2|...\leq |\beta_{2M}|$.
Solving $|\beta_M|=|\beta_{M+1}|$, one finds that $\beta$ traces out a closed loop $C_\beta$ on the complex plan referred to as the 
generalized Brillouin zone. 
Applying this result to our nonreciprocal stub model with $M=1$, Eq. \eqref{eign-H-beta} is a quadratic equation for $\beta$ and has two solutions
$\beta_{1,2}$. The continuum condition requires $\left|\beta_{1}\right|=\left|\beta_{2}\right|$ and leads to
\begin{equation}
|\beta_{1,2}| = r = \left| \frac{ t_{1}-\frac{\kappa}{2} }{ t_{1}+\frac{\kappa}{2}}\right|^{\frac{1}{2}}
\label{eq:betas}.
\end{equation}
Thus the generalized Brillouin zone is a circle with radius $r$. In the Hermitian limit $\kappa=0$,
$C_\beta$ reduces to the unit circle. One can check that for $\beta\in C_\beta$, the eigenvalues of $H_\beta$
indeed form continuum bands  that match the open chain spectrum such as the one shown in 
Fig. \ref{fig:chain} (except for the edge states at $\pm t_3$).

The introduction of generalized Brillouin zone enables us to derive the correct topological phase transition point where
the edge states appear/vanish. In the generalized band theory based on $H(\beta)$, the transition at $t_1=t_c$ corresponds to the point
where the continuum band touches the edge state energy. 
Setting $E=t_3$ in Eq. \eqref{eign-H-beta}, we find 
\begin{equation}
\beta_1=-\frac{t_2}{t_1+\frac{\kappa}{2}},\;\;\; \beta_2=-\frac{t_1-\frac{\kappa}{2}}{t_2}.
\end{equation}
Requiring  $\left|\beta_{1}\right|=\left|\beta_{2}\right|$ yields
\begin{align}
   & t_{1}=\pm\sqrt{t_{2}^{2}+\left(\frac{\kappa}{2}\right)^{2}}&\quad\text{for}\quad\frac{\kappa}{2}<t_{1},\label{subeq:tca}\\
    &t_{1}=\pm\sqrt{-t_{2}^{2}+\left(\frac{\kappa}{2}\right)^{2}}&\quad\text{for}\quad\frac{\kappa}{2}>t_{1}.\label{subeq:tcb}
\end{align}
which prove Eqs. \eqref{tca} and \eqref{tcb} earlier. Note that $t_{c}$ here does not coincide with any band gap closing as in the SSH model. Closing of the band gap is not a prerequisite for topological transitions, even in Hermitian systems \cite{Ezawa13}. 
As we will show below in Secs. \ref{sec:majorana} and \ref{sec:sqrt}, at the transition point $t_c$, the topological invariant undergoes 
a jump and becomes ill-defined, illustrating the topological origin of the edges states.

In Refs. \cite{Yao18,Yoko19},  the bulk-boundary correspondence is reestablished by introducing a quantized winding number $w$ 
along $C_\beta$. This is not possible in the present case due to the lack of chiral symmetry. To see this, we define the Zak phase \cite{Zak89}
for the $m$-th band as follows:
\begin{equation}
\gamma_m = i \oint_{C_\beta} d\beta \langle \lambda_m(\beta)|\partial_\beta |\psi_m(\beta)\rangle,
\end{equation}
where $|\psi_m(\beta)\rangle$ and $\langle \lambda_m(\beta)|$ are the right and left eigenvector of $H(\beta)$ 
corresponding to the $m$-th eigenenergy $\{E_m(\beta)\}$. Note that a non-Hermitian Hamiltonian has both left and right eigenstates defined as $H^{\dagger}\ket{\lambda_{m}}=E_m^{*}\ket{\lambda_{m}}$ and $H\ket{\psi_{m}}=E\ket{\psi_{m}}$. Compared to the Hermitian case, the set of eigenstates $\left\{ \ket{\psi_{n}}\right\}$ are not necessarily orthogonal but are linearly independent \cite{Brody14}. 
We follow the convention of biorthogonal normalization, $\braket{\lambda_{i}|\psi_{j}}=\delta_{i,j}$.
In the limit of $t_3=0$, the stub model reduces to the
nonreciprocal SSH model, the Zak phase for the two bands with finite Re$E$ is then quantized to multiples of $\pi$:
$\pi$ for $|t_1|<\sqrt{t_2^2+(\kappa/2)^2}$ and zero otherwise.
For finite $t_3$, they are no longer multiples of $\pi$. This can be illustrated for example by considering the limit $t_1=0$ and 
$\kappa=0$, the Zak phase for the positive energy band is $-\pi t_2^2/(t^2_2+t_3^2)$.
Thus $\gamma_m$ cannot serve as the topological invariants for the stub model. 

\section{\label{sec:majorana}Invariant from Majorana Stars}

A convenient way to visualize the eigenstates of a multiband Bloch Hamiltonian is to represent 
them as a set of stars on the Bloch sphere through Majorana's stellar representation \cite{majorana1932,bruno2012,niu2012quantum,Liu14,Teo20,Xu20}. More specifically, the eigenvectors of $H(\beta)$ with $(2j+1)$ bands
can be viewed as $\beta$-dependent spinors of spin $j$,
\begin{equation}
\ket{\psi_{m}\left(\beta\right)}=\big[C^{(m)}_{j},C^{(m)}_{j-1},\ldots,C^{(m)}_{-j}\big]^{T},
\end{equation}
where $m$ is the band index. 
Recall that any spin-$j$ state $\ket{\Phi}$ can be constructed using two bosonic creation operators $a_{\uparrow,\downarrow}^{\dagger}$ following Schwinger's bosonic representation of angular momentum eigenstates \cite{Schwinger52}:
\begin{equation}
\ket{\Phi}=\frac{1}{N_{j}}\prod_{\ell=1}^{2j}\big[\cos \frac{\theta_{\ell}}{2} a_{\uparrow}^{\dagger}+\sin \frac{\theta_{\ell}}{2} e^{i\phi_{\ell}}a_{\downarrow}^{\dagger}\big]\ket{0}
\label{eq:Schwinger},
\end{equation}
where $N_j$ is the normalization factor and $\ket{0}$ is the vacuum state. 
The parametrization in Eq. \eqref{eq:Schwinger} makes it clear that 
the spin-$j$ state is represented by $2j$ points living on the Bloch sphere, 
labelled by index $\ell$ with polar angle $\theta_{\ell}$ and azimuthal angle $\phi_{\ell}$ respectively. 
Each point on the Bloch sphere can be viewed as a spin-1/2 state
generated by acting $\big[\cos \frac{\theta_{\ell}}{2} a_{\uparrow}^{\dagger}+\sin \frac{\theta_{\ell}}{2} e^{i\phi_{\ell}}a_{\downarrow}^{\dagger}\big]$ on the vacuum. We will refer to these points as Majorana stars. Together they form a ``quantum constellation,''
which encodes the band topologies.  

To find the positions of the stars for a given state such as $\ket{\psi_{m}\left(\beta\right)}$,
it is sufficient to solve for the roots of the so-called star equation, a polynomial equation of 
degree $2j$ for complex variable $z$  \cite{majorana1932},
\begin{equation}
\sum_{k=0}^{2j}\frac{\left(-1\right)^{k}C^{(m)}_{j-k}}{\sqrt{\left(2j-k\right)!k!}}z^{2j-k}=0.
\label{eq:MSR}
\end{equation}
Once the $2j$ roots $\{z^{(m)}_{\ell}\}$ are obtained, the angular positions of the stars can be determined by 
\begin{equation}
z^{(m)}_{\ell}=\tan\frac{\theta^{(m)}_{\ell}\left(\beta\right)}{2}e^{i\phi^{(m)}_{\ell}\left(\beta\right)},\;\;\ell=1,2,...2j,
\end{equation}
where we have restored the $\beta$ dependence of the angles.

\begin{figure}[htpb!]
\includegraphics[width=0.45\textwidth]{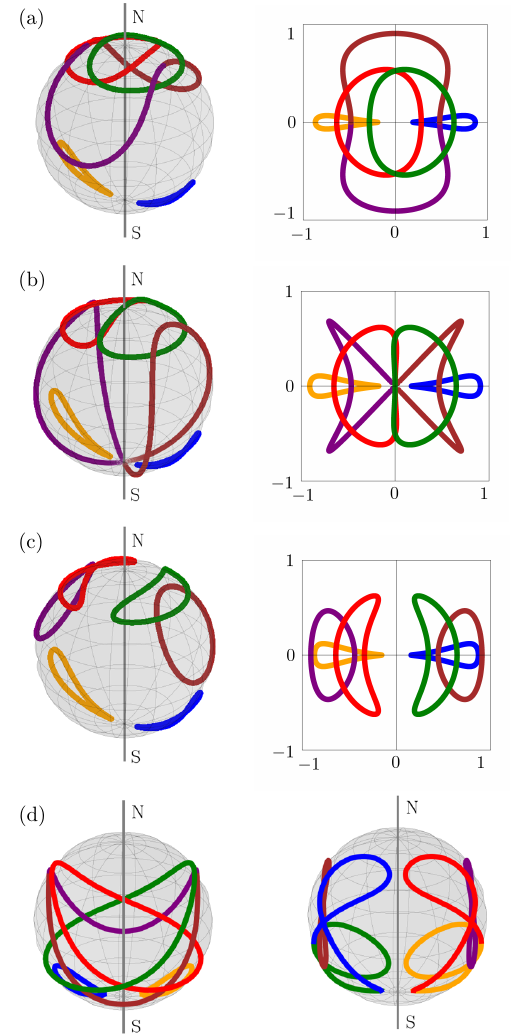}
\caption{\label{fig:majoranastars} Majorana stars (left panel) and their projections on the $xy$ plane (right) for (a) $t_{1}=1.1$ inside the topologically nontrivial phase; (b) $t_{1}=t_c$ at the transition point; and (c) $t_1=1.3$ inside the trivial phase. The red and orange (blue and green) curves are the trajectories of Majorana stars for the 
band with Re$E>0$ (Re$E<0$). The brown and purple curves correspond to the band at zero energy. The stars are constructed from the right eigenvectors of $H(\beta)$ for $\beta$ within the generalized Brillouin zone. (d) Majorana stars obtained from the left eigenvectors of $H(\beta)$ for $t_{1}=1.1$ (left) and $1.3$ (right).  Parameters used are $t_{2}=1$, $t_{3}=0.25$, and $\kappa=4/3$.}
\end{figure}

Our nonreciprocal stub model has three bands and corresponds to a spin-1 system, $j=1$. 
Thus, each band is represented by two Majorana stars on the Bloch sphere. 
As $\beta$ is varied throughout the generalized Brillouin zone $C_\beta$,
the pair of stars for a given band trace out closed curves on the Bloch sphere as 
shown in Figs. \ref{fig:majoranastars}(a) to \ref{fig:majoranastars}(c). Here for the sake of clarity,
the star trajectories are also projected onto the $xy$ plane and depicted in panels on the right.
In Hermitian systems, the solid angles subtended  by the closed trajectories of the Majorana stars are 
related to the Zak phase \cite{Liu14,Liu16}. Here for the stub model, the Zak phase is not quantized 
and the solid angles of the closed star trajectories vary smoothly with parameters
such as $t_1$. Yet, the azimuthal winding number of the star trajectories is always an integer.
We find that the azimuthal winding contains enough information to distinguish the topological nontrivial phase 
(with edge states) from the trivial phase. 

More specifically, the azimuthal winding number for the $m$-th band is defined as
\begin{equation}
\nu_{m}=-\frac{1}{2\pi}\sum_{\ell=1}^{2j}\oint_{C_\beta} d\beta \partial_{\beta}\phi^{(m)}_{\ell}
\label{eq:MSWind}.
\end{equation}
The winding is easy to count by visual inspection. For example, for the band with Re$E>0$, the red (orange)
curve in Fig. \ref{fig:majoranastars}(a) winds around the $z$ axis one (zero) time, so the total winding
is 1. Similarly, for the  Re$E<0$ band, the green (blue) curve winds one (zero) time. And finally, for the band
at zero energy, the brown and purple curve together contribute to winding number 1. To summarize, in this case of $t_1$=1.1,
all three bands have $\nu_m=1$. 
In comparison, for the case of $t_1=1.3$ shown in Fig. \ref{fig:majoranastars}(c),
none of the curves wind around the $z$ axis, $\nu_m=0$, as is evident from their projection on the $xy$ plane. 

The winding number $\nu_m$ defined in Eq. \eqref{eq:MSWind} depends on the choice of gauge or basis. For example, 
in a new basis where the site order is switched from $(b,a,c)$ to $(a,b,c)$, only the zero-energy band has winding number 1. 
This example also shows that the total winding $\nu_{total}=\sum_m\nu_m$ is gauge-dependent, but
$\nu_{total}=3$ and $\nu_{total}=1$ are equivalent. 
In fact, the parity of the total winding number is gauge invariant,
\begin{equation}
P=(-1)^{\sum_m \nu_{m}}.
\label{eq:MSParity}
\end{equation}
Then we are naturally led to the following definition: phases with $P=-1$ ($P=1$), i.e. with $\nu_{total}$ odd (even),
are topologically nontrivial (trivial).
The topological invariant $P$ introduced here based on Majorana stars correctly predicts the phase diagram and the phase transition
point. For example, Fig. \ref{fig:majoranastars}(a) has $P=-1$ and depicts a topologically nontrivial phase, while $P=1$ 
in Fig. \ref{fig:majoranastars}(c). At the transition point $t_1=t_c$ depicted in Fig. \ref{fig:majoranastars}(b), 
the red and green curve pass through the north pole, while the brown and purple line pass through the south pole,
at which point the corresponding azimuthal winding number is ill-defined. 

\begin{figure}
\includegraphics[width=0.45\textwidth]{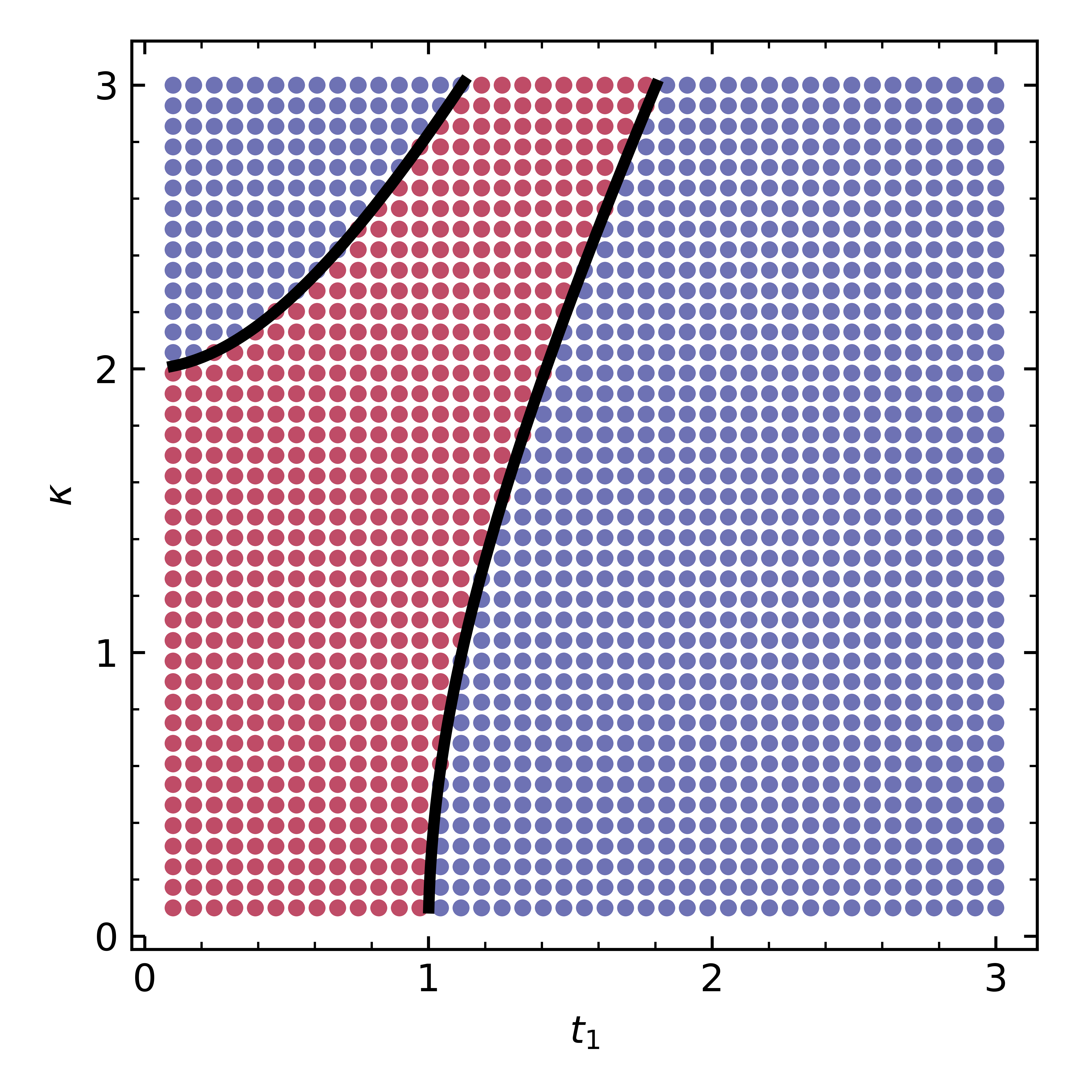}
\caption{\label{fig:phase} The phase diagram for the nonreciprocal stub model with parameters $t_{2}=1$ and $t_{3}=0.25$. 
The topological invariant $P$ defined in Eq. \eqref{eq:MSParity} is computed numerically. The topological nontrivial (trivial) region is shown in red (blue) and has a value $P=-1$ ($P=1$).  The upper left phase boundary is given by Eq. \eqref{subeq:tcb} while the lower right phase boundary is given by Eq. \eqref{subeq:tca}.}
\end{figure}

In summary, by analytically continuing $H(k)$ to $H(\beta)$
and introducing the topological invariant $P$ based on the Majorana star representation, we have re-established the bulk-boundary correspondence
for the nonreciprocal stub model. This is confirmed by comparing the phase diagram predicted from the numerical evaluation of 
invariant $P$, shown in Fig. \ref{fig:phase}, with the edge state spectra of open chains. The red region with $P=-1$ features edge states, while
the blue region with $P=1$ does not. The numerical phase diagram also confirms the analytical phase boundaries given by Eqs. \eqref{subeq:tca} and 
 \eqref{subeq:tcb} shown in solid lines. Other parameters of $t_2$ and $t_3$ can be discussed in a similar fashion.

In the formulation above we have exclusively relied on the right eigenstates. Alternatively, we can introduce Majorana stars
and the winding numbers based on the left eigenstates of $H(\beta)$. Two examples are shown in Fig. \ref{fig:majoranastars}(d).
Compared to their respective counterpart obtained from the right eigenstates in Figs. \ref{fig:majoranastars}(a) and \ref{fig:majoranastars}(c), the
star trajectories appear very different, but the winding number $\nu_m$ retains the same magnitude with the sign flipped.

\section{\label{sec:sqrt}Parent Hamiltonian}

To gain further insight about the band topology, we now view the stub model from another angle by analyzing it's parent Hamiltonian which
is considerably simpler. This perspective is inspired by one-dimensional square-root topological insulators ($\sqrt{\mathrm{TI}}$).
A $\sqrt{\mathrm{TI}}$ described by a Hamiltonian $H$ is an insulator whose topological properties are inherited from its parent Hamiltonian $H^{2}$ \cite{Arkinstall17,Ezawa20}. While $H$ does not fit into the standard description of topological insulators, $H^2$ does. 
The first clue that the stub model is potentially a square-root topological insulator is the existence of in-gap edge states away from zero energy, a common feature of $\sqrt{\mathrm{TI}}$s. The second indication is that a similar, but different, three-band model, the Hermitian diamond lattice threaded with $\phi$ flux, has been recently identified as a $\sqrt{\mathrm{TI}}$ \cite{Kremer20}. The diamond lattice model shares a few common features with the stub model here including the presence of zero energy flat band and nonquantized Zak phase. In what follows, we show that the nonreciprocal stub model is not a 
$\sqrt{\mathrm{TI}}$ as defined in recent works. Its parent Hamiltonian $H^2$ does not possess chiral symmetry or quantized Zak phase, and therefore is not in the same league of the SSH model. Despite this, the parent Hamiltonian helps elucidate what happens to the azimuthal winding across the phase transition point $t_1=t_c$.

The parent Hamiltonian for the nonreciprocal stub model is a function of $\beta$
\begin{equation}
H^{2}(\beta)=\left[\begin{array}{ccc}
m_{0}(\beta)-t_{3}^{2} & 0 & m_{-}+t_{2}t_{3}\beta\\
0 & m_{0}(\beta) & 0\\
m_{+}+t_{2}t_{3}\beta^{-1} & 0 & t_{3}^{2}
\end{array}\right]
\label{eq:HamSquared}
\end{equation}
where $m_{0}(\beta)=(t_{1}+t_{2}\beta-\frac{\kappa}{2} ) (t_{1}+t_{2}\beta^{-1}+\frac{\kappa}{2})+t_{3}^{2}$ and 
$m_{\pm}=\left(t_{1}\pm\frac{\kappa}{2}\right)t_{3}$. 
Recall the child Hamiltonian $H$ has three bands, $E(\beta)=0,\pm\sqrt{m_{0}}(\beta)$. 
The eigenvalues of the parent Hamiltonian $H^{2}$ consist of only two bands, $\mathcal{E}(\beta)=0$ and $m_{0}(\beta)$, since
the two original bands of $H$ with opposite Re$E$ become degenerate after the square.

The form of $H^2$ in Eq. \eqref{eq:HamSquared} is not very convenient. We can transform it into 
block diagonal form, which is always possible
because the original stub model lives on a bipartite lattice \cite{Marques21}. 
This is simply achieved by a reordering of the basis to
\begin{equation}
H^{2}=H_{res}\oplus H_{saw}.
\end{equation} 
Now the parent Hamiltonian is reduced to the direct sum of a 
residual Hamiltonian 
\begin{equation}
H_{res} (\beta )=m_{0}(\beta)
\end{equation}
which describes hopping along a chain, and a $2\times 2$ Hamiltonian
\begin{equation}
H_{saw} (\beta)=\left[\begin{array}{cc}
m_{0}(\beta)-t_{3}^{2}& m_{-}+t_{2}t_{3}\beta\\
m_{+}+t_{2}t_{3}\beta^{-1} & t_{3}^{2}
\end{array}\right]
\label{eq:HamTI}
\end{equation}
which can be interpreted as a tight-binding model with nearest neighbor hopping on the sawtooth lattice. 
The generalized Brillouin zone for $H_{saw}$ coincides with that of $H$, and the 
two bands are $0$ and $m_0(\beta)$. Note that most hopping models on the sawtooth lattice do not feature a flat band
except for certain special ratios of the hopping amplitudes. The reason why $H_{saw}$ hosts a zero-energy flat band is because 
there are also onsite potential terms $m_{0}-t_{3}^{2}$ and $t^2_3$. 

\begin{figure}[htpb!]
\includegraphics[width=0.45\textwidth]{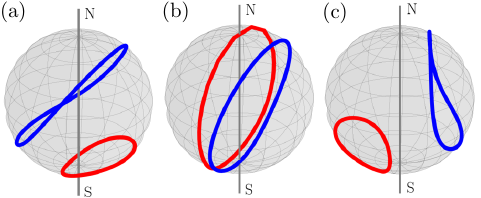}
\caption{\label{fig:H2wind} The right eigenvectors of $H_{saw}(\beta)$ on the Bloch sphere as $\beta$ is varied across the generalized Brillouin zone.  The red (blue) curve corresponds to the band at zero ($m_0$). The values for $t_{1}$ are (a) $t_{1}=1.1$; (b) $t_{1}=t_c=1.2$; and $t_1=1.3$ with
$t_{2}=1$, $t_{3}=0.25$, and $\kappa=4/3$.
}
\end{figure}

As a two band model, the right eigenvectors of $H_{saw}(\beta)$ can be brought into the standard form $\ket{\psi(\beta)}=\cos \frac{\theta}{2}\ket{0}+\sin \frac{\theta}{2} e^{i\phi}\ket{1}$ to live on the Bloch sphere. We stress that $H_{saw}$ contains all three Pauli components, $H_{saw}=\mathcal{E}_0+\mathbf{d}\cdot\boldsymbol{\sigma}$, so its eigenvectors in general do not span a great circle.
As $\beta$ is varied throughout the generalized Brillouin zone $C_\beta$, the eigenvector of each band traces out a closed curve on the Bloch sphere. Fig. \ref{fig:H2wind} shows the trajectory of the zero (in red) and $m_0$ band (in blue) for $t_1$ values before, at, and after the transition point $t_1=t_c$. The azimuthal winding of these curves is reminiscent of the Majorana stars, only to show up more clearly thanks to the reduction in the number of bands. Define the winding of azimuthal angles for each band as
\begin{equation}
\nu=-\frac{1}{2\pi}\oint d\beta \partial_{\beta}\phi. 
\label{eq:SquaredWind}
\end{equation}
Note that the two bands have the same winding number so it is sufficient to focus on one of them, say the $m_0$ band.
This can be seen for example in the Hermitian limit, where $H_{saw}$ can be shifted, flattened, and cast into the form $\hat{n}(k)\cdot \boldsymbol{\sigma}$. Then the two eigenvectors are antipodal points on the Bloch sphere and wind around the $z$ axis in the same way.
We find that for $t\in [0,t_c)$, $\nu=1$ for the  $m_0$ bands, the red and blue curves both enclose the $z$-axis
as shown in Fig. \ref{fig:H2wind}(a). In contrast, for $t>t_c$, $\nu=0$, neither 
the red nor the blue trajectory winds around the $z$-axis, see Fig. \ref{fig:H2wind}(c). Right at the transition point $t=t_c$ [Fig. \ref{fig:H2wind}(b)], the red curve crosses the north pole while the blue curve
crosses the south pole, at which points $\phi$ becomes ill defined. 
The winding number is quantized and jumps by one at the transition.
The transition is topological in the sense that it is impossible to smoothly vary the blue curve in Fig. \ref{fig:H2wind}(a) to that in Fig. \ref{fig:H2wind}(c) without going thought one of the poles. 

\section{\label{sec:sum}Conclusion}

The nonreciprocal stub lattice model appears deceivingly simple. Yet understanding its bulk-edge correspondence 
is not a straightforward matter and requires concepts and techniques developed only recently. 
Its bulk spectrum has a flat band and two exceptional points separate  phases characterized by distinct knots of the eigenenergy strings. 
The bulk phase transition points however do not coincide with the emergence of edge states in finite systems with open boundaries.
This failure of the traditional (Hermitian) bulk-boundary correspondence is accompanied by the NH skin effect. For finite chains with open boundaries, 
the continuum bands away from zero energy all congregate to one edge, but the flat band resists the skin effect and has the lowest degree 
of localization as measured by the IPR. A NH bulk-boundary correspondence is established by analytically continuing 
the Hamiltonian $H(k)$ to $H(\beta)$ with the complex $\beta$ confined within the generalized Brillouin zone $C_\beta$.
The resultant continuum band structure
gives correct prediction of the critical point $t_c$ where the edge states onset. 

The three-band model $H(\beta)$ differs from the NH SSH model or its generalization in one important aspect, it does not possess a chiral symmetry so the Zak phase is not quantized to $0$ or $\pi$ (or fractions of $\pi$ such as $\pi/2$). It also differs from known examples of $\sqrt{\mathrm{TI}}$s, because its parent Hamiltonian, or more precisely its subblock $H_{saw}$, does not feature a quantized Zak phase and cannot be identified as a known topological insulator.  Despite those differences, $H^2(\beta)$ can still be utilized to accurately characterize the NH band topology. This suggests that parent Hamiltonian is useful beyond traditional $\sqrt{\mathrm{TI}}$s. By representing the eigenvectors of  $H(\beta)$ using Majorana stars, we find that at the transition, the Majorana star trajectories pass through the north or south pole, triggering a jump in the azimuthal winding number. We propose that the parity $P$ of the total azimuthal winding can serve as the gauge-independent invariant to characterize different gapped phases. Phases with odd parity have edge states. And it is impossible to go from an odd parity phase to an even parity phase without having the star trajectory crossing the poles. Using the invariant $P$, one can reliably predict the existence/absence of edge state from the bulk information of $H(\beta)$. Note that the parity does not give the total number of isolated edge states.

A challenging open question is to formulate a rigorous proof of the bulk-edge relationship based on the Majorana stellar representation. Previous studies for Hermitian systems \cite{bruno2012,Liu14} have shown that the geometrical phases of MSs are closely related to the Zak phases of the system and the existence of edge states is associated with the nontrivial winding of MSs \cite{Mong11}. Generalizing the proof to non-Hermitian systems, including establishing the exact correspondence between $P$ and the appearance of topological edge states found here, is an open challenge left for future work. We speculate that for generic multiband NH systems, the Zak phases of the individual bands only provide partial topological data, while the Majorana stars contain the full information.  Future work is required to fully understand the  topological information contained in the quantum constellation and their experimental signatures in edge spectrum and quantum dynamics. We hope the analysis presented here can be extended to the topological characterizations of other multi-band models.

The nonreciprocal stub lattice can be realized experimentally using electric circuits \cite{Zhao18,Lee18} or optical ring resonators \cite{Longhi15}. 
For example, electric circuits consisting of LC resonators with negative impedance converters can achieve nonreciprocal hopping, and 
admittance measurement can probe the complex band structures of NH lattice modes \cite{Helbig20}. Similarly, ring resonators can realize the nonreciprocal lattice by using antiresonant coupling rings to produce asymmetric hopping \cite{Longhi15}. The presence of edge states can then be demonstrated by measuring transmittance and imaging the propagation of light \cite{Mittal14}.

\acknowledgments
This work is supported by AFOSR Grant No. FA9550- 16-1-0006 and NSF Grant No. PHY- 2011386.

\bibliography{stubbibfile}

\end{document}